\documentclass[11pt,preprint,showpacs,preprintnumbers,amsmath,amssymb]{revtex4}

\usepackage{graphicx}
\usepackage{dcolumn}
\usepackage{bm}
\usepackage{multirow}
\usepackage{graphicx}
\usepackage{float}
\usepackage{subfigure}
\usepackage{diagbox}
\usepackage{CJK}
\usepackage{caption}
\usepackage{color}

\begin{document}

\title{Analyzing $\Xi(1620)$ in the molecule picture in the Bethe-Salpeter equation approach}

\author{Zhen-Yang Wang \footnote{e-mail: wangzhenyang@nbu.edu.cn}}
\affiliation{\scriptsize{Physics Department, Ningbo University, Zhejiang 315211, China}}

\author{Jing-Juan Qi \footnote{e-mail: qijj@mail.bnu.edu.cn}}
\affiliation{\scriptsize{College of Nuclear Science and Technology, Beijing Normal University, Beijing 100875, China}}

\author{Jing Xu \footnote{e-mail: xj2012@mail.bnu.edu.cn}}
\affiliation{\scriptsize{Physics Department, Yantai University, Yantai 264005, China}}

\author{Xin-Heng Guo \footnote{Corresponding author, e-mail: xhguo@bnu.edu.cn}}
\affiliation{\scriptsize{College of Nuclear Science and Technology, Beijing Normal University, Beijing 100875, China}}

\date{\today}

\begin{abstract}
In this work, we assume that the observed state $\Xi(1620)$ is a $s$-wave $\Lambda\bar{K}$ or $\Sigma\bar{K}$ bound state. Based on this molecule picture, we establish the Bethe-Salpeter equations for $\Xi(1620)$  in the ladder and instantaneous approximations. We solve the Bethe-Salpeter equations for the $\Lambda\bar{K}$ and $\Sigma\bar{K}$ systems numerically and find that the $\Xi(1620)$ can be explained as $\Lambda\bar{K}$ and $\Sigma\bar{K}$ bound states with $J^P=1/2^-$, respectively. Then we calculate the decay widths of $\Xi(1620)\rightarrow\Xi\pi$ in these two different molecule pictures systems, respectively.

\end{abstract}


\maketitle
\section{Introduction}
The advent of the LHCb, Belle, BES$\mathrm{\uppercase\expandafter{\romannumeral3}}$, and other facilities and their unexpectedly successful contributions to hadron physics have stimulated of hadron studies. With the observations of some states which do not agree well with the theoretical predictions in the constituent quark model (like $\Lambda(1405)$, $\Xi(1620)$, $X$, $Y$, $Z$ states, and pentaquark states ($P_c(4380)$ and $P_c(4450)$ states)) \cite{Tanabashi:2018oca}, it is important to study these unusual states, both to probe the limitations of the quark model and to discover the unrevealed aspects of the quantum chromodynamics (QCD) description of structures of hadron resonances.

Up to now lots of nucleons and $S = \pm1$ hyperon resonances have been discovered and their quantum numbers have also been measured. In the charmed baryon sector, there have been also lots of significant progresses made in the experimental studies by the LHCb, Belle, BES$\mathrm{\uppercase\expandafter{\romannumeral3}}$ and other collaborations. For $\Xi$ states, only the spin-parity quantum numbers of the ground octet state $\Xi(1320)$, the decuplet state $\Xi(1530)$, and the excited state $\Xi(1820)$ have been determined, but for other known $\Xi$ resonances, their spin-parity numbers are incomplete. Foe example, the $\Xi(1690)$ and $\Xi(1620)$ states are cataloged in the Particle Data Group (PDG) with only one-star and three-star \cite{Tanabashi:2018oca}, respectively. If $\Xi(1620)$ has $J^P$ as $1/2^-$, it will be similar to the $\Lambda(1405)$ state, which has been postulated as a meson-baryon molecular state or a pentaquark candidate \cite{Hyodo:2011ur}. Determining the masses and quantum numbers of the $\Xi$ resonances is vital for us to understand their structures.

The $\Xi(1620)$ was observed through the $\Xi(1620)\rightarrow \Xi\pi$ decay in the 1970's \cite{Ross:1972bf,Briefel:1977bp}. Although the mass and the width measurements in the two experiments are consistent, they both have large statistical uncertainties. Recently, the Belle Collaboration reported the observation of $\Xi(1620)$ via its decay to $\Xi^-\pi^+$ happened in the $\Xi^+_c\rightarrow\Xi^-\pi^+\pi^+$ decay \cite{Sumihama:2018moz}. The mass and width are measured to be $1610.4\pm6.0(\mathrm{stat})^{+5.9}_{-3.5}(\mathrm{syst})$ MeV and $59.9\pm4.8(\mathrm{stat})^{+2.8}_{-3.0}(\mathrm{syst})$ MeV, respectively.

On the theoretical side, one was shown that it is very difficult to accommodate the $\Xi(1620)$ in the quark models \cite{Capstick:1986bm,Blask:1990ez}. On the other hand, the meson-baryon scattering in the strangeness $S = -2$ sector was also studied in different unitary coupled-channel approaches constrained by QCD chiral symmetry \cite{Ramos:2002xh,GarciaRecio:2003ks,Miyahara:2016yyh}. In all these chiral unitary approaches, the $\Xi(1620)$ is dynamically generated with a relatively large decay width, and couples strongly to the $\Xi\pi$ and $\Lambda\bar{K}$ channels but very weakly to $\Sigma\bar{K}$ and $\Xi\eta$. In addition, the poles of $\Xi(1620)$ are below the threshold of $\Lambda\bar{K}$.

The purpose of this paper is to study the possibilities that the $\Xi(1620)$ is a $\Lambda\bar{K}$ or $\Sigma\bar{K}$ bound state with quantum numbers $J^P$ = $1/2^-$ in the Bethe-Salpeter equation approach. We will also calculate the decay widths of $\Xi(1620)\rightarrow\Xi\pi$ in these two pictures. The Bethe-Salpeter equation is a formally exact equation to describe the relativistic bound state \cite{Salpeter:1951sz,lurie-book}, and
has been applied in many theoretical studies concerning heavy mesons and heavy baryons \cite{Guo:1996jj,Guo:2007mm,Guo:1998ef,Wang:2018jaj,Wang:2017smo,Wang:2017dcq}. In this paper, we will study the $s$-wave baryon-meson molecular bound state with the kernel introduced by the vector meson exchange interactions.

This paper is organized as follows. In the next section, we will briefly review the Bethe-Salpeter equation for the bound state of a meson and a baryon. In Sec. \ref{normalization-condition-se}, we will discuss the normalization condition of the Bethe-Salpeter wave function. In Sec. \ref{decay-se}, the decay of $\Xi(1620)\rightarrow\Xi\pi$ will be calculated. The numerical results will be presented in Sec. \ref{sec-result}. In the last section, we will give a summary.

\section{the bethe-salpeter formalism for the $\Xi(1620)$}
\label{sect-BS}
In this section, we will review the general formalism of the Bethe-Salpeter equation and derive the Bethe-Salpeter equation for the system composed of a baryon ($\Lambda$ or $\Sigma$) and a pseudoscalar meson ($\bar{K}$). Then we will derive the normalization condition for the Bethe-Salpeter wave function in the next section. Firstly, we define the Bethe-Salpeter wave function for the bound state $|P\rangle$ of a baryon ($\Lambda$ or $\Sigma$) and a pseudoscalar meson ($\bar{K}$) as the following:
\begin{equation}
  \chi\left(x_1,x_2,P\right) = \langle0|T\psi(x_1)\phi(x_2)|P\rangle,
\end{equation}
where $\psi(x_1)$ and $\phi(x_2)$ are the field operators of the baryon ($\Lambda$ or $\Sigma$) and pseudoscalar meson ($\bar{K}$) at space coordinates $x_1$ and $x_2$, respectively, $P$ denotes the total momentum of the bound state with mass $M$ and velocity $v$. In momentum space, the Bethe-Salpeter wave function can be defined as
\begin{equation}\label{momentum-BS-function}
 \chi_P(x_1,x_2,P) = e^{-iPX}\int\frac{d^4p}{(2\pi)^4}e^{-ipx}\chi_P(p),
\end{equation}
where $p$ represents the relative momentum of the two constituents.

The Bethe-Salpeter equation for the bound state can be written in the following form:
\begin{equation}\label{BS-equation}
  \chi_{P}(p)=S_\psi(p_1)\int\frac{d^4q}{(2\pi)^4}K(P,p,q)\chi_{P}(q)S_{\bar{\phi}}(p_2),
\end{equation}
where $S_\psi(p_1)$ and $S_{\bar{\phi}}(p_2)$ are the propagators of the baryon ($\Lambda$ or $\Sigma$) and the pseudoscalar meson ($\bar{K}$), respectively, and $K(P,p,q)$ is the kernel which contains two-particle-irreducible diagrams. For convenience, we define $p_l (=p\cdot v)$ and $p_t^\mu(=p^\mu- p_lv^\mu)$ to be the longitudinal and transverse  projections of the relative momentum ($p$) along the bound state momentum ($P$). Then, the propagator of $\Lambda$ ( or $\Sigma$) has the form
\begin{equation}\label{baryon-propagator}
  S_\psi(\lambda_1P+p)=\frac{i\left[\left(\lambda_1M+p_l\right)v\!\!\!/+p\!\!\!/_t+m_1\right]}{\left(\lambda_1M+p_l+\omega_1-i\epsilon\right)\left(\lambda_1M+p_l-\omega_1+i\epsilon\right)}.
\end{equation}
and the propagator of the $\bar{K}$ meson can be expressed as
\begin{equation}\label{pseudoscalar-propagator}
  S_{\bar{K}}(\lambda_2P-p)=\frac{i}{(\lambda_2M-p_l+\omega_2-i\epsilon)(\lambda_2M-p_l-\omega_2+i\epsilon)},
\end{equation}
where $\omega_{1(2)}=\sqrt{m_{1(2)}^2+p_t^2}$ (in which we have defined $p_t^2=-p_t\cdot p_t$), $\lambda_1 = m_1/(m_1 + m_2)$ and $\lambda_2 = m_2/(m_1 + m_2)$, $m_1$ and $m_2$ are the masses of $\Lambda(\Sigma)$ and $K$ mesons

In general, for a baryon and a pseudoscalar meson bound state, considering $v\!\!\!/u(v,s) = u(v,s)$ ($u(v,s)$ is the spinor of the bound state with helicity $s$), $\chi_{P}(p)$ can be written as
\begin{equation}
  \chi_{P}(p) = \left(g_1+g_2\gamma_5+g_3\gamma_5p\!\!\!/_t + g_4p\!\!\!/_t + g_5\sigma_{\mu\nu}\varepsilon^{\mu\nu\alpha\beta}p_{t\alpha}v_\beta\right),
\end{equation}
where $g_i$ ($i$ = 1, $\cdot\cdot\cdot$, 5) are Lorentz-scalar functions. Furthermore, each term in the expansion of $\chi_{P}(p)$ transforms exactly in the way that $\chi_{P}(p)$ transforms under $P$-parity
and Lorentz transformations, which can help us simplify the form of $\chi_{P}(p)$, it is easy to prove that $\chi_{P}(p)$ can be simplified as
\begin{equation}\label{BS-function}
  \chi_{P}(p) = [f_1(p)+f_2(p)p\!\!\!/_t]u(v,s),
\end{equation}
in which $f_1(p)$ and $f_2(p)$ are two independent Lorentz-scalar function of $p$.

As discussed in the introduction, we will study the $s$-wave bound state of the $\Lambda \bar{K}$ and $\Sigma \bar{K}$ systems. The isospin field doublets $\psi = \left(\psi^0,\psi^-\right)^T$ and $\phi = \left(-\phi^+,\phi^0\right)^T$ have the following expansions in momentum space:
\begin{equation}
  \begin{split}
    \psi_1(x) &= \int\frac{d^3p}{(2\pi)^3\sqrt{2E_\psi^{\pm}}}\left(a_{\psi^-}e^{-ipx}+a_{\psi^+}^\dag e^{ipx}\right), \\
    \psi_2(x) &= \int\frac{d^3p}{(2\pi)^3\sqrt{2E_\psi^0}}\left(a_{\psi^0}e^{-ipx}+a_{\bar{\psi}^0}^\dag e^{ipx}\right), \\
    \phi_1(x) &= \int\frac{d^3p}{(2\pi)^3\sqrt{2E_\phi^{\pm}}}\left(a_{\phi^+}e^{-ipx}+a_{\phi^-}^\dag e^{ipx}\right), \\
    \phi_2(x) &= \int\frac{d^3p}{(2\pi)^3\sqrt{2E_\phi^0}}\left(a_{\phi^0}e^{-ipx}+a_{\bar{\phi}^0}^\dag e^{ipx}\right). \\
  \end{split}
\end{equation}

The isospin quantum number of $\Xi(1620)$ is 1/2, so the flavor wave function of $\Lambda \bar{K}$ and $\Sigma \bar{K}$ systems can be written as
\begin{equation}
  |P\rangle_{\frac12,\frac12} =|\Lambda^0\bar{K}^0 \rangle,
\end{equation}
\begin{equation}
  |P\rangle_{\frac12,\frac12} =\sqrt{\frac{2}{3}}|\Sigma^+K^-\rangle -  \frac{1}{\sqrt{3}}|\Sigma^0K^0 \rangle.
\end{equation}

Projecting the bound states on the field operators $\psi_1(x)$, $\psi_2(x)$, $\phi_1(x)$, and $\phi_2(x)$, then we have
\begin{equation}
  \langle0|T{\psi_i(x_1)\phi_j(x_2)}|P\rangle_{I,I_3}=C_{(I,I_3)}^{ij}\chi_P^{I}\left(x_1,x_2\right),
\end{equation}
where $\chi_P^{I}$ is the common Bethe-Salpeter wave function for the bound state with isospin $I$. The isospin coefficient $C_{(I,I_3)}^{ij}$ is
\begin{equation}
  C_{(\frac12,-\frac12)}^{22} = 1,
\end{equation}
for the $\Lambda \bar{K}$ system, and the isospin coefficients are
\begin{equation}
  C_{(\frac12,-\frac12)}^{11} = \sqrt{\frac{2}{3}}, \quad C_{(\frac12,-\frac12)}^{22} = -\sqrt{\frac{1}{3}},
\end{equation}
for the $\Sigma \bar{K}$ system.

Then the Bethe-Salpeter equation for the bound state can be written as
\begin{equation}
  C_{(I,I_3)}^{ij}\chi_P^{I}(p) = S_\psi(\lambda_1P+p) \int\frac{d^4q}{(2\pi)^4}K^{ij,lk}\left(P,p,q\right) C_{(I,I_3)}^{lk}\chi_P^{I}(q)S_{\phi}(\lambda_2P-p),
\end{equation}
where $i(j)$ and $l(k)$ refer to the components of the $\psi(\phi)$ field doublets. Then, the Bethe-Salpeter equation for the $I$ = 1/2 $\Lambda\bar{K}$ molecule can be written as
\begin{equation}
  \chi_P(p) = S_\Lambda(\lambda_1P+p) \int\frac{d^4q}{(2\pi)^4}K^{22,22}\chi_P(q)S_{\bar{K}}(\lambda_2P-p),
\end{equation}
and for the $I$ = 1/2 $\Sigma\bar{K}$ molecule the Bethe-Salpeter equation can be write as
\begin{equation}
  \chi_P(p) = S_\Sigma(\lambda_1P+p) \int\frac{d^4q}{(2\pi)^4}\left(K^{11,11}-\frac{1}{\sqrt{2}}K^{11,22}\right)\chi_P(q)S_{\bar{K}}(\lambda_2P-p).
\end{equation}

In the Bethe-Salpeter equation approach, the interactions between $\Lambda$ and $\bar{K}$ mesons are due to the light vector-meson ($\omega$ and $\phi$) exchanges. There is no $\rho$-exchange contribution, because of the isospin conservation. For the $\Sigma \bar{K}$ interaction we will consider the exchanges of vector mesons $\rho$, $\omega$ and $\phi$. The pseudoscalar meson exchanges are forbidden because the $K$ meson is also a pseudoscalar meson. The Lagrangians for the vertices of the strange $K$ meson and one-strange baryon with vector mesons are \cite{Ronchen:2012eg,He:2017aps}:
\begin{equation}
\begin{split}
\mathcal{L}_{KK\rho}&=ig_{KK\rho}\bar{K}\rho^\mu\cdot\tau\partial_{\mu}K+c.c.,\\
\mathcal{L}_{KK\omega}&=ig_{KK\omega}\bar{K}\omega^\mu\partial_{\mu}K+c.c.,\\
\mathcal{L}_{KK\phi}&=ig_{KK\phi}\bar{K}\phi^\mu\partial_{\phi}K+c.c.,\\
\mathcal{L}_{BB\rho}&=-g_{BB\rho}\bar{B}\left[\gamma^\nu-\frac{\kappa_{BB\rho}}{2m_{B}}B^{\nu\rho}\partial_{\rho}\right]\boldsymbol{\rho}_\nu \cdot\boldsymbol{\tau}B,\\
\mathcal{L}_{BB\omega}&=-g_{BB\omega}\bar{B}\left[\gamma^\nu-\frac{\kappa_{BB\omega}}{2m_{B}}\sigma^{\nu\rho}\partial_{\rho}\right]\omega_\nu B,\\
\mathcal{L}_{BB\phi}&=-g_{BB\phi}\bar{B}\left[\gamma^\nu-\frac{\kappa_{BB\phi}}{2m_{\Sigma}}B^{\nu\rho}\partial_{\rho}\right]\phi_\nu B,\\
\end{split}
\end{equation}
where $c.c.$ is the complex conjugate of the first term, $\boldsymbol{\tau}$ is the Pauli spin matrix. The coupling constants are constrained by the $SU(3)$ symmetry, $g_{KK\rho}=g_{KK\omega}=g_{\rho\pi\pi}/2$ and $g_{KK\phi}=g_{\rho\pi\pi}/\sqrt{2}$. The $\rho\pi\pi$ coupling is determined by $g_{\rho\pi\pi}=M_{\rho}/(\sqrt{2}f_\pi)\approx 6.1$. $g_{\Lambda\Lambda\omega}=\frac23g_{NN\rho}(5\alpha-2)$ and $g_{\Lambda\Lambda\phi}=-\frac{\sqrt{2}}{3}g_{NN\rho}(2\alpha+1)$, where we take the value $\alpha = 1.15$  based on the $\omega$ coupling constant given in Ref. \cite{Janssen:1996kx}. $g_{\Sigma\Sigma\rho}=g_{\Sigma\Sigma\omega}=2\alpha g_{NN\rho}$ and $g_{\Sigma\Sigma\phi}=-\sqrt{2}(2\alpha-1)g_{NN\rho}$. $g_{NN\rho}$ is chosen as $g_{\rho\pi\pi}/2$ as in Ref.\cite{Ronchen:2012eg,He:2017aps},  Under $SU(3)$ symmetry, the $\kappa_{BBV} (B= \Lambda, \Sigma)$ can be obtained with the relations $f_{\Lambda\Lambda\omega}=\frac56 f_{NN\omega}-\frac12f_{NN\rho}$, $f_{\Lambda\Lambda\phi}=-\frac{1}{3\sqrt{2}}$, $f_{NN\omega}-\frac{1}{\sqrt{2}}f_{NN\rho}$, $f_{\Sigma\Sigma\rho}=f_{\Sigma\Sigma\omega}=(f_{NN\omega}+f_{NN\rho})/2$, and $f_{\Sigma\Sigma\phi}=(-f_{NN\omega}+f_{NN\rho})/\sqrt{2}$, where $f_{BBV}$ is defined as $f_{BB\rho}=g_{BB\rho}\kappa_{BB\rho}$, and $\kappa_{BB\rho}=6.1$ and $f_{NN\omega}=0$ \cite{Ronchen:2012eg}.

From the above observations, at the tree level, in the $t$-channel we have the following kernel for the Bethe-Salpeter equation in the so-called ladder approximation:
\begin{equation}
K(P,p,q)=c_Ig_{\Sigma\Sigma V}g_{KKV}\left(\gamma^\alpha+\frac{i\kappa_{\Sigma\Sigma V}}{2m_{\Sigma}}\sigma^{\alpha\beta}q_{1\beta}\right)(p_2+q_2)^\mu\Delta_{\alpha\mu}(k,m_V),
\end{equation}
where $m_V$ represents the mass of the exchanged vector meson ($\rho$, $\omega$ and $\phi$), and $c_I$ is the isospin coefficient: $c_{1/2}$ = $1-\sqrt{2}$, 1, 1 for $\rho$, $\omega$, $\phi$ mesons, respectively.

In order to describe the phenomena in the real world, we should include a form factor at each interacting vertex of hadrons to include the finite-size effects of these hadrons. For the meson-exchange case, the form factor is assumed to take the following form \cite{Lohse:1990ew}:
\begin{equation}\label{form-factor}
F(k)=\frac{\Lambda^2-m^2}{\Lambda^2-k^2},
\end{equation}
where $\Lambda$, $m$ and $k$ represent the cutoff parameter, the mass of the exchanged meson and the momentum of the exchanged meson, respectively.

Substituting Eqs. (\ref{baryon-propagator}), (\ref{pseudoscalar-propagator}), (\ref{BS-function}), and (\ref{form-factor}) into Eq. (\ref{BS-equation}) and using the so-called covariant instantaneous approximation \cite{Guo:1996jj}, $p_l=q_l$, we obtain
\begin{equation}
\begin{split}
[f_1(p)+f_2(p)p\!\!\!/_t]=&\frac{ic_Ig_{\Sigma\Sigma V}g_{KKV}[(\lambda_1M+p_l)v\!\!\!/+p\!\!\!/_t+m_1]}{[(\lambda_1M+p_l)^2-\omega_1^2+i\epsilon][(\lambda_2M-p_l)^2-\omega_2^2+i\epsilon]}\int\frac{d^4q}{(2\pi)^4}\\
&\frac{2(\lambda_2M-p_l)v\!\!\!/-p\!\!\!/_t-q\!\!\!/_t-(p\!\!\!/_t-q\!\!\!/_t)(p_t^2-q_t^2)/m_V^2}{-(p_t-q_t)^2-m_V^2}F^2(k,m_V)[f_1(q)+f_2(q)q\!\!\!/_t].
\end{split}
\end{equation}
Then we obtain the following coupled integral equations for $f_1(p)$ and $f_2(p)$:
\begin{equation}\label{4-D-f1-wavefunction}
\begin{split}
f_1(p)&=\frac{ig_{\Sigma\Sigma V}g_{KKV}}{(\lambda_1M+p_l+\omega_1-i\epsilon)(\lambda_1M+p_l-\omega_1+i\epsilon)(\lambda_2M-p_l+\omega_2-i\epsilon)(\lambda_2M-p_l-\omega_2+i\epsilon)}\\
&\int\frac{d^4q}{(2\pi)^4}\Bigg\{\frac{2(\lambda_1M+p_l)(\lambda_2M-p_l)+p_t^2+p_t\cdot q_t+(p_t^2-p_t\cdot q_t)(p_t^2-q_t^2)/m_V^2}{-(p_t-q_t)^2-m_V^2}f_1(q)\\
&+\frac{m_1[p_t\cdot q_t+q_t^2+(p_t\cdot q_t-q_t^2)(p_t^2-q_t^2)/m_V^2]}{-(p_t-q_t)^2-m_V^2}f_2(q)\Bigg\}F^2(k,m_V),
\end{split}
\end{equation}

\begin{equation}\label{4-D-f2-wavefunction}
\begin{split}
f_2(p)p_t^2&=\frac{-ig_{\Sigma\Sigma V}g_{KKV}}{(\lambda_1M+p_l+\omega_1-i\epsilon)(\lambda_1M+p_l-\omega_1+i\epsilon)(\lambda_2M-p_l+\omega_2-i\epsilon)(\lambda_2M-p_l-\omega_2+i\epsilon)}\\
&\int\frac{d^4q}{(2\pi)^4}\Bigg\{\frac{m_1[p_t^2+p_t\cdot q_t+(p_t^2-p_t\cdot q_t)(p_t^2-q_t^2)/m_V^2]}{-(p_t-q_t)^2-m_V^2}f_1(q)\\
&+\frac{-2(\lambda_1M+p_l)(\lambda_2M-p_l)p_t\cdot q_t-p_t^2[p_t\cdot q_t+q_t^2+(p_t\cdot q_t -q_t^2)(p_t^2-q_t^2)/m_V^2]}{-(p_t-q_t)^2-m_V^2}f_2(q)\Bigg\}F^2(k,m_V).
\end{split}
\end{equation}
We notice that in Eqs. (\ref{4-D-f1-wavefunction}) and (\ref{4-D-f2-wavefunction}) there are poles in $p_l$ at $-\lambda_1M-\omega_1+i\epsilon$, $-\lambda_1M+\omega_1-i\epsilon$, $\lambda_2M+\omega_2-i\epsilon$ and $\lambda_2M-\omega_2+i\epsilon$. By choosing the appropriate contour, we integrate over $p_l$ on both sides of Eqs. (\ref{4-D-f1-wavefunction}) and (\ref{4-D-f2-wavefunction}) and obtain the following coupled integral equations for $\tilde{f}_1(p_t)$ and $\tilde{f}_2(p_t)$

\begin{equation}\label{3-D-f1-wavefunction}
\begin{split}
\tilde{f}_1(p_t)=&\frac{g_{\Sigma\Sigma V}g_{KKV}}{2\omega_1(M+\omega_1+\omega_2)(M+\omega_1-\omega_2)}\int\frac{d^3q_t}{(2\pi)^3}\Bigg[\frac{p_t\cdot q_t+q_t^2+(p_t\cdot q_t-q_t^2)(p_t^2-q_t^2)/m_V^2}{-(p_t-q_t)^2-m_V^2}m_1\tilde{f}_2(q_t)\\
&+\frac{-2\omega_1(M+\omega_1)+p_t^2+p_t\cdot q_t+(p_t^2-p_t\cdot q_t)(p_t^2-q_t^2)/m_V^2}{-(p_t-q_t)^2-m_V^2}\tilde{f}_1(q_t)\Bigg]F^2(k_t)\\
&-\frac{g_{\Sigma\Sigma V}g_{KKV}}{2\omega_2(M+\omega_1-\omega_2)(M-\omega_1-\omega_2)}\int\frac{d^3q_t}{(2\pi)^3}\Bigg[\frac{p_t\cdot q_t+q_t^2+(p_t\cdot q_t-q_t^2)(p_t^2-q_t^2)/m_V^2}{-(p_t-q_t)^2-m_V^2}m_1\tilde{f}_2(q_t)\\
&+\frac{2\omega_2(M-\omega_2)+p_t^2+p_t\cdot q_t+(p_t^2-p_t\cdot q_t)(p_t^2-q_t^2)/m_V^2}{-(p_t-q_t)^2-m_V^2}\tilde{f}_1(q_t)\Bigg]F^2(k_t),
\end{split}
\end{equation}

\begin{equation}\label{3-D-f2-wavefunction}
\begin{split}
\tilde{f}_2(p_t)p_t^2=&\frac{-g_{\Sigma\Sigma V}g_{KKV}}{2\omega_1(M+\omega_1+\omega_2)(M+\omega_1-\omega_2)}\int\frac{d^3q_t}{(2\pi)^3}\Bigg[\frac{p_t^2+p_t\cdot q_t+(p_t^2-p_t\cdot q_t)(p_t^2-q_t^2)/m_V^2}{-(p_t-q_t)^2-m_V^2}m_1\tilde{f}_1(q_t)\\
&+\frac{2\omega_1(M+\omega_1)p_t\cdot q_t-p_t^2[p_t\cdot q_t+q_t^2+(p_t\cdot q_t- q_t^2)(p_t^2-q_t^2)/m_V^2]}{-(p_t-q_t)^2-m_V^2}\tilde{f}_2(q_t)\Bigg]F^2(k_t)\\
&+\frac{g_{\Sigma\Sigma V}g_{KKV}}{2\omega_2(M+\omega_1-\omega_2)(M-\omega_1-\omega_2)}\int\frac{d^3q_t}{(2\pi)^3}\Bigg[\frac{p_t^2+p_t\cdot q_t+(p_t^2-p_t\cdot q_t)(p_t^2-q_t^2)/m_V^2}{-(p_t-q_t)^2-m_V^2}m_1\tilde{f}_1(q_t)\\
&+\frac{-2\omega_2(M-\omega_2)p_t\cdot q_t-p_t^2[p_t\cdot q_t+q_t^2+(p_t\cdot q_t- q_t^2)(p_t^2-q_t^2)/m_V^2]}{-(p_t-q_t)^2-m_V^2}\tilde{f}_2(q_t)\Bigg]F^2(k_t),
\end{split}
\end{equation}
where $\tilde{f}_{1(2)}(p_t)\equiv\int dp_lf_{1(2)}(p)$.

After reducing the above coupled integral equations for $\tilde{f}_1(p_t)$ and $\tilde{f}_2(p_t)$ to one dimensional integral equations, we obtain the following equations:
\begin{equation}\label{one-demention-BS}
\begin{split}
\tilde{f}_1(|p_t|)&=A_{11}(|p_t|,|q_t|)\tilde{f}_1(|q_t|)+A_{12}(|p_t|,|q_t|)\tilde{f}_2(|q_t|),\\
\tilde{f}_2(|p_t|)&=A_{21}(|p_t|,|q_t|)\tilde{f}_1(|q_t|)+A_{22}(|p_t|,|q_t|)\tilde{f}_2(|q_t|),\\
\end{split}
\end{equation}
where $A_{ij}(|p_t|,|q_t|)$ $(i,j=1,2)$ are of the following forms:
\begin{equation}
\begin{split}
A_{11}(|p_t|,|q_t|)&=\frac{-g_{\Sigma\Sigma V}g_{KKV}|q_t|}{8m_V^2|p_t|\omega_1\omega_2(M+\omega_1-\omega_2)[M^2-(\omega_1+\omega_2)^2]}\Bigg\{\frac{4|p_t||q_t|(\Lambda^2-m_V^2)}{[\Lambda^2+(|p_t|-|q_t|)^2][\Lambda^2+(|p_t|+|q_t|)^2]}\\
&\times\bigg\{8\omega_1\omega_2M^2m_V^2+\Lambda^2(m_V^2-|p_t|^2+|q_t|^2)[M(\omega_1-\omega_2)+(\omega_1+\omega_2)^2]\\
&+(\omega_1+\omega_2)^2[(|p_t|-|q_t|)^2+m_V^2(3|p_t|^2+|q_t|^2-4\omega_1\omega_2)]\\
&+M(\omega_1-\omega_2)[(|p_t|^2-|q_t|^2)^2+m_V^2(3|p_t|^2+|q_t|^2+4\omega_1\omega_2)]\bigg\}\\
&+\bigg\{8\omega_1\omega_2M^2m_V^2+(\omega_1+\omega_2)^2[m_V^4+(|p_t|^2-|q_t|^2)^2+2m_V^2(p_t^2+q_t^2-2\omega_1\omega_2)]\\
&+M(\omega_1-\omega_2)[m_V^4+(|p_t|^2-|q_t|^2)^2+2m_V^2(p_t^2+q_t^2-2\omega_1\omega_2)]\bigg\}\ln\frac{\Lambda^2+(|p_t|+|q_t|)^2}{\Lambda^2+(|p_t|-|q_t|)^2}\\
&-\bigg\{8\omega_1\omega_2M^2m_V^2+(\omega_1+\omega_2)^2[m_V^4+(|p_t|^2-|q_t|^2)^2+2m_V^2(p_t^2+q_t^2-2\omega_1\omega_2)]\\
&+M(\omega_1-\omega_2)[m_V^4+(|p_t|^2-|q_t|^2)^2+2m_V^2(p_t^2+q_t^2-2\omega_1\omega_2)]\bigg\}\ln\frac{m_V^2+(|p_t|+|q_t|)^2}{m_V^2+(|p_t|+|q_t|)^2}\Bigg\},\\
\end{split}
\end{equation}
\begin{equation}
\begin{split}
A_{12}(|p_t|,|q_t|)&=\frac{-g_{\Sigma\Sigma V}g_{KKV}m_1|q_t|}{8m_V^2|p_t|\omega_1\omega_2(M+\omega_1-\omega_2)[M^2-(\omega_1+\omega_2)^2]}\Bigg\{\frac{4|p_t||q_t|(\Lambda^2-m_V^2)}{[\Lambda^2+(|p_t|-|q_t|)^2][\Lambda^2+(|p_t|+|q_t|)^2]}\\
&+[m_V^4+(|p_t|^2-|q_t|^2)^2+2m_V^2(|p_t|^2+|q_t|^2)]\ln\frac{\Lambda^2+(|p_t|+|q_t|)^2}{\Lambda^2+(|p_t|-|q_t|)^2}\\
&-[m_V^4+(|p_t|^2-|q_t|^2)^2+2m_V^2(|p_t|^2+|q_t|^2)]\ln\frac{m_V^2+(|p_t|+|q_t|)^2}{m_V^2+(|p_t|-|q_t|)^2}\Bigg\},\\
\end{split}
\end{equation}
\begin{equation}
\begin{split}
A_{21}(|p_t|,|q_t|)&=\frac{-g_{\Sigma\Sigma V}g_{KKV}m_1|q_t|[M(\omega_1-\omega_2)+(\omega_1+\omega_2)^2]}{8m_V^2|p_t|^3\omega_1\omega_2(M+\omega_1-\omega_2)[M^2-(\omega_1+\omega_2)^2]}\\
&\Bigg\{\frac{4|p_t||q_t|(\Lambda^2-m_V^2)[(|p_t|^2-|q_t|^2)^2+\Lambda^2(m_V^2-|p_t|^2+|q_t|^2)+m_V^2(2|p_t|+|q_t|^2)]}{[\Lambda^2+(|p_t|-|q_t|)^2][\Lambda^2+(|p_t|+|q_t|)^2]}\\
&-[m_V^4+(|p_t|^2-|q_t|^2)^2+2m_V^2(|p_t|^2+|q_t|^2)]\ln\frac{\Lambda^2+(|p_t|+|q_t|)^2}{\Lambda^2+(|p_t|-|q_t|)^2}\\
&+[m_V^4+(|p_t|^2-|q_t|^2)^2+2m_V^2(|p_t|^2+|q_t|^2)]\ln\frac{m_V^2+(|p_t|+|q_t|)^2}{m_V^2+(|p_t|-|q_t|)^2}\Bigg\},\\
\end{split}
\end{equation}
\begin{equation}
\begin{split}
A_{22}(|p_t|,|q_t|)&=\frac{-g_{\Sigma\Sigma V}g_{KKV}|q_t|}{8m_V^2|p_t|^3\omega_1\omega_2(M+\omega_1-\omega_2)[M^2-(\omega_1+\omega_2)^2]}\Bigg\{\frac{4|p_t||q_t|(\Lambda^2-m_V^2)}{[\Lambda^2+(|p_t|-|q_t|)^2][\Lambda^2+(|p_t|+|q_t|)^2]}\\
&\bigg\{4\omega_1\omega_2M^2m_V^2(|p_t|^2+|q_t|^2)+4\omega_1\omega_2\Lambda^2M^2m_V^2\\
&+(\omega_1+\omega_2)^2[(|p_t|^3-|p_t||q_t|^2)^2+m_V^2(|p_t|^4-2\omega_1\omega_2|p_t|^2-2\omega_1\omega_2|q_t|^2+3|p_t|^2|q_t|^2)]\\
&+\Lambda^2(\omega_1+\omega_2)^2[|p_t|^4-|p_t|^2|q_t|^2+m_V^2(|p_t|^2-2\omega_1\omega_2)]\\
&+M\Lambda^2(\omega_1-\omega_2)[|p_t|^4-|p_t|^2|q_t|^2+m_V^2(|p_t|^2+2\omega_1\omega_2)]\\
&+M(\omega_1-\omega_2)[(|p_t|^3-|p_t||q_t|^2)^2+m_V^2(|p_t|^4+2\omega_1\omega_2|p_t|^2+2\omega_1\omega_2|q_t|^2+3|p_t|^2|q_t|^2)]\bigg\}\\
&+\bigg\{(\omega_1+\omega_2)^2[(|p_t|^3-|p_t||q_t|^2)^2+m_V^4(|p_t|^2-2\omega_1\omega_2)+2m_V^2(|p_t|^2+|q_t|^2)(|p_t|^2-\omega_1\omega_2)]\\
&+M\omega_1-\omega_2[(|p_t|^3-|p_t||q_t|^2)^2+2m_V^2(|p_t|^2+|q_t|^2)(|p_t|^2+\omega_1\omega_2)+m_V^4(|p_t|^2+\omega_1\omega_2)]\\
&+4\omega_1\omega_2M^2m_V^2(m_V^2+|p_t|^2+|q_t|^2)\bigg\}\ln\frac{\Lambda^2+(|p_t|+|q_t|)^2}{\Lambda^2+(|p_t|-|q_t|)^2}\\
&-\bigg\{(\omega_1+\omega_2)^2[(|p_t|^3-|p_t||q_t|^2)^2+m_V^4(|p_t|^2-2\omega_1\omega_2)+2m_V^2(|p_t|^2+|q_t|^2)(|p_t|^2-\omega_1\omega_2)]\\
&+M\omega_1-\omega_2[(|p_t|^3-|p_t||q_t|^2)^2+2m_V^2(|p_t|^2+|q_t|^2)(|p_t|^2+\omega_1\omega_2)+m_V^4(|p_t|^2+\omega_1\omega_2)]\\
&+4\omega_1\omega_2M^2m_V^2(m_V^2+|p_t|^2+|q_t|^2)\bigg\}\ln\frac{\Lambda^2+(|p_t|+|q_t|)^2}{\Lambda^2+(|p_t|-|q_t|)^2}.\\
\end{split}
\end{equation}

\section{The normalization condition for the bound state}
\label{normalization-condition-se}

The normalization condition for a baryon and a pseudoscalar meson bound state is given by \cite{lurie-book}
\begin{equation}\label{normalization}
\frac{i}{(2\pi)^4}\int d^4pd^4 \bar{\chi}_P(p)\frac{\partial}{\partial P_0}\left[I(p,q,P)+K(p,q,P)\right]\chi_P(q)=2P_0,
\end{equation}
where $I(p,q,P)$ is the inverse of the four-point propagator defined as follows:
\begin{equation}
I(p,q,P)=\delta^{(4)}(p-q)[S_\Sigma(\lambda_1P+p)]^{-1}[S_K(\lambda_2P-p)]^{-1}.
\end{equation}

After some algebra, the normalization condition in Eq. (\ref{normalization}) can be written in the following form as in Refs. \cite{Weng:2010rb,Liu:2015qfa,Wang:2017rjs}:
\begin{equation}\label{C-normalization-condition}
\begin{split}
-\int\frac{d^4p}{(2\pi)^4}&\big\{\mathrm{Tr}[\alpha_P(p)\beta_P(p)S_\Sigma(p_1)(\lambda_1\varepsilon\!\!\!/)S_\Sigma(p_1)S_K(p_2)]\\
&+\mathrm{Tr}[\alpha_P(p)\beta_P(p)(2\lambda_2p_2\cdot\varepsilon)S_\Sigma(p_1)S_K(p_2)S_K(p_2)]\big\}=2P_0,
\end{split}
\end{equation}
where $\varepsilon=(1,\vec{0})$, $\alpha_P(p)$ and $\beta_P(p)$ are the transverse projections of the Bethe-Salpeter wave functions given as follows:
\begin{equation}
\begin{split}
\alpha_P(p)&=-iS_\Sigma(p_1)^{-1}\chi_P(p)S_K(p_2)^{-1},\\
\beta_P(p)&=-iS_K(p_2)^{-1}\bar{\chi}_P(p)S_\Sigma(p_2)^{-1}.\\
\end{split}
\end{equation}

Substituting Eq. (\ref{BS-equation}) into  above equations, then, one can derive the parametric forms of $\alpha_P(p)$ and $\beta_P(p)$ as
\begin{equation}\label{alpha-beta}
\begin{split}
\alpha_P(p)&=[\tilde{h}_1(p_t)+p\!\!\!/_t\tilde{h}_2(p_t)]u(v,s),\\
\beta_P(p)&=\bar{u}(v,s)[\tilde{h}_1(p_t)+p\!\!\!/_t\tilde{h}_2(p_t)],\\
\end{split}
\end{equation}
with
\begin{equation}\label{h1-h2}
\begin{split}
\tilde{h}_1(p_t)&=\int \frac{d^3q_t}{(2\pi)^3}\frac{p_t^2+p_t\cdot q_t+(p_t^2-p_t\cdot q_t)(p_t^2-q_t^2)/m_V^2}{(p_t-q_t)^2+m_V^2}\tilde{f}_2(q_t),\\
\tilde{h}_2(p_t)&=-\int \frac{d^3q_t}{(2\pi)^3}\frac{p_t^2+p_t\cdot q_t+(p_t^2-p_t\cdot q_t)(p_t^2-q_t^2)/m_V^2}{p_t^2[(p_t-q_t)^2+m_V^2]}\tilde{f}_1(q_t).
\end{split}
\end{equation}

After substituting Eqs. (\ref{alpha-beta}) and (\ref{h1-h2}) into Eq. (\ref{C-normalization-condition}), we have
\begin{equation}
\begin{split}
&i\int\frac{d^4p}{(2\pi)^4}\bigg\{\Big\{\tilde{h}_1^2(p_t)\lambda_1(\lambda_1M+p_l)[(\lambda_1M+p_l)^2-p_t^2+3m_1^2]-6\tilde{h}_1(p_t)\tilde{h}_2(p_t)p_t^2m_1\lambda_1(\lambda_1M+p_l)\\
&-\tilde{h}_2^2(p_t)p_t^2\lambda_1(\lambda_1M+p_l)[(\lambda_1M+p_l)^2-p_t^2+3m_1^2]\Big\}/\Big\{2m_1[(\lambda_1M+p_l)^2-\omega_1^2]^2[(\lambda_2M-p_l)^2-\omega_2^2]\Big\} \\
&+2\lambda_2(\lambda_2M-p_l)\Big\{\tilde{h}_1^2(p_t)[(\lambda_1M+p_l)^2-p_t^2+m_1^2]-4\tilde{h}_1(p_t)\tilde{h}_2(p_t)p_t^2m_1 \\
&-\tilde{h}_2^2(p_t)p_t^2[(\lambda_1M+p_l)^2-p_t^2+m_1^2]\Big\}/\Big\{2m_1[(\lambda_1M+p_l)^2-\omega_1^2][(\lambda_2M-p_l)^2-\omega_2^2]^2\Big\}\bigg\}=2P_0.\\
\end{split}
\end{equation}

\section{Numerical results for the bethe-salpeter wave functions}
\label{sec-result}

In this part, we will solve the Bethe-Salpeter equation numerically and try to search for possible solutions of the $\Lambda\bar{K}$ and $\Sigma\bar{K}$ bound states. It can be seen from Eq. (\ref{one-demention-BS}) that there is only one free parameter in our model, the cutoff $\Lambda$, which contains the information about the nonpoint interactions due to the structure of hadrons at the interaction vertices. Although the value of $\Lambda$ cannot be exactly determined and depends on the specific process, it should be typically the scale of low energy physics, which is about 1 GeV. In this work, we treat the cutoff in the form factors as a parameter varying in a much wider range $0.8-4.8$ GeV.

To find out the possible molecule bound states, one only needs to solve the homogeneous Bethe-Salpeter equations.  One numerical solution of the homogeneous Bethe-Salpeter equation corresponds to a possible bound state.  The integration region in each integral will be discretized into $n$ pieces, with $n$ being sufficiently large. In this way, the integral equation will be converted into an $n\times n$ nmatrix equation, and the scalar wave functions of each equation will now be regarded as an $n$-dimensional vector. Then, the two coupled integral equations can be illustrated as

\begin{equation}
\left(
\begin{matrix}
 \tilde{f}_1(|p_t|)\\
 \tilde{f}_2(|p_t|)\\
\end{matrix}
\right)=\left(
\begin{matrix}
A_{11}(|p_t|,|q_t|)    & A_{12}(|p_t|,|q_t|)  \\
A_{21}(|p_t|,|q_t| )   & A_{21} (|p_t|,|q_t|)\\
\end{matrix}
\right)
\left(\begin{matrix}
 \tilde{f}_1(|q_t|)\\
 \tilde{f}_2(|q_t|)
\end{matrix}
\right),
\end{equation}
where $\tilde{f}_{1(2)}$ is an $n$-dimensional vector, and $A_{ij}(|p_t|,|q_t|)(i,j=1,2)$ is an $n \times n$ matrix,
which corresponds to the matrix labeled by $p_t$ and $q_t$ in each integral equation. Generally, $|p_t|$ (and $|q_t|$) varies from 0 to $+\infty$. Here, $|p_t|$ (and $|q_t|$) will be transformed into a new variable $t$ that varies from $-1$ to 1 based on the Gaussian integration method,
\begin{equation}
|p_t|=\epsilon+w\log\left[1+y\frac{1+t}{1-t}\right],
\end{equation}
where $\epsilon$ is a parameter introduced to avoid divergence in numerical calculations, $w$ and $y$ are
parameters used in controlling the slope of wave functions and finding the proper solutions
for these functions. Then one can obtain the numerical results of the Bethe-Salpeter wave functions by
requiring the eigenvalue of the eigenvalue equation to be 1.

\begin{figure}[htbp]
\centering
\subfigure[\ The Lorentz-scalar function of $\tilde{f}_1(|p_t|)$]{
\begin{minipage}[t]{0.45\linewidth}
\centering
\includegraphics[width=3.05in]{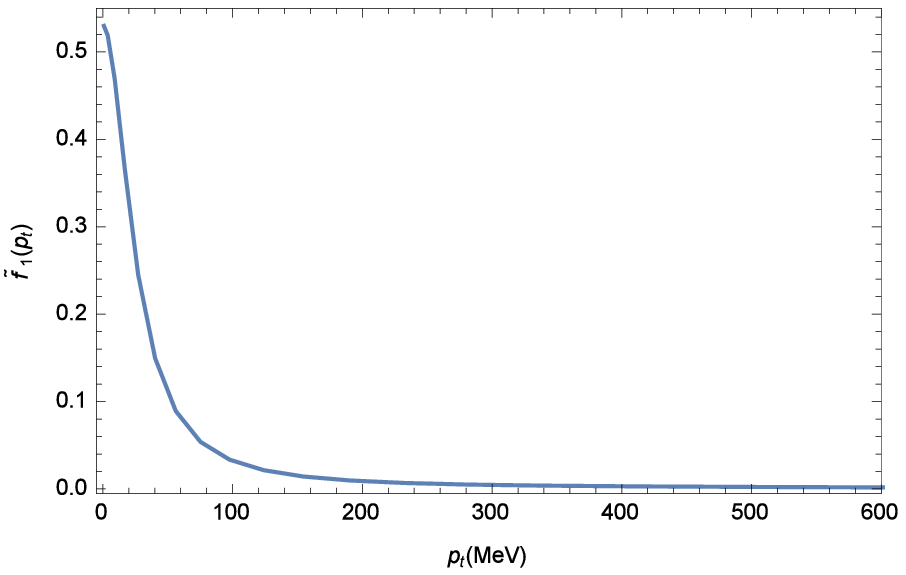}
\end{minipage}%
}%
\subfigure[\ The Lorentz-scalar function of $\tilde{f}_2(|p_t|)$]{
\begin{minipage}[t]{0.5\linewidth}
\centering
\includegraphics[width=3.3in]{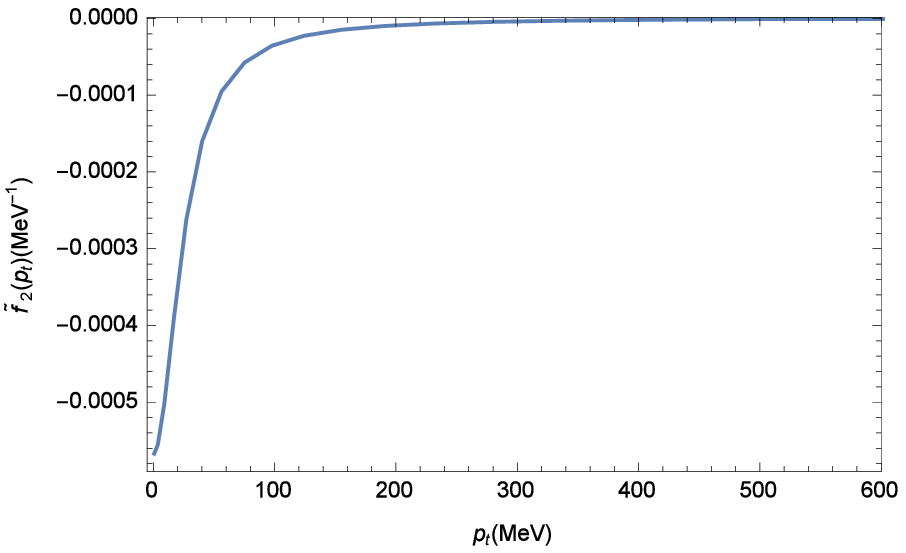}
\end{minipage}
}%
\centering
\caption{Numerical results for the Bethe-Salpeter wave functions in the $\Lambda\bar{K}$ system.}
\label{lambdaK}
\end{figure}

\begin{figure}[htbp]
\centering
\subfigure[\ The Lorentz-scalar function of $\tilde{f}_1(|p_t|)$.]{
\begin{minipage}[t]{0.45\linewidth}
\centering
\includegraphics[width=3.12in]{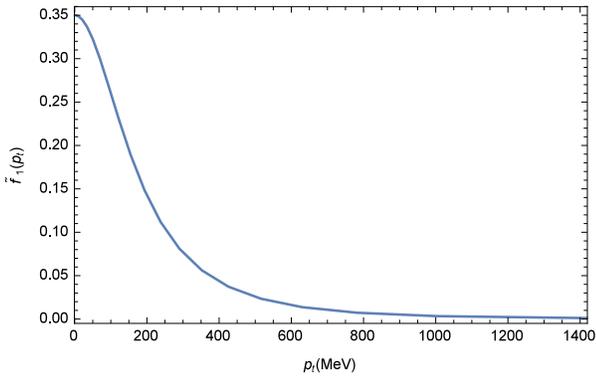}
\end{minipage}%
}%
\subfigure[\ The Lorentz-scalar function of $\tilde{f}_2(|p_t|)$.]{
\begin{minipage}[t]{0.5\linewidth}
\centering
\includegraphics[width=3.3in]{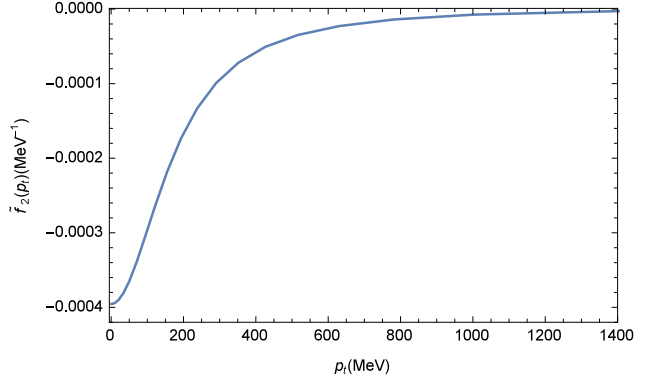}
\end{minipage}
}%
\centering
\caption{Numerical result for the Bethe-Salpeter wave functions in the $\Sigma\bar{K}$ system.}
\label{SigmaK}
\end{figure}

In our calculation, we take the masses of the mesons and baryons from the PDG \cite{Tanabashi:2018oca,Sumihama:2018moz}, $m_{\Xi(1620)}=1610.4$ MeV, $m_{\Lambda}=1115.683$ MeV, $m_{\Sigma}= 1187.354$ MeV, $m_{\Xi}= 1314.86$ MeV, $m_{K}= 494.988$ MeV $m_{\pi}= 139.571$ MeV. From our calculations, we find $\Lambda\bar{K}$ and $\Sigma\bar{K}$ systems can be $\Xi(1620)$ state when the cutoff $\Lambda$ = 1632 MeV and 1356 MeV, respectively. The corresponding numerical results of the Lorentz-scalar functions in the normalized Bethe-Salpeter equation, $\tilde{f}_1(|p_t|)$ and $\tilde{f}_2(|p_t|)$, are given in Figs. \ref{lambdaK} and \ref{SigmaK} for the $\Lambda\bar{K}$ and $\Sigma\bar{K}$ systems, respectively

 \section{the decay of $\Xi(1620)\rightarrow\Xi\pi$}
\label{decay-se}

\begin{figure}[ht]
\centering
    \rotatebox{0}{\includegraphics*[width=0.50\textwidth]{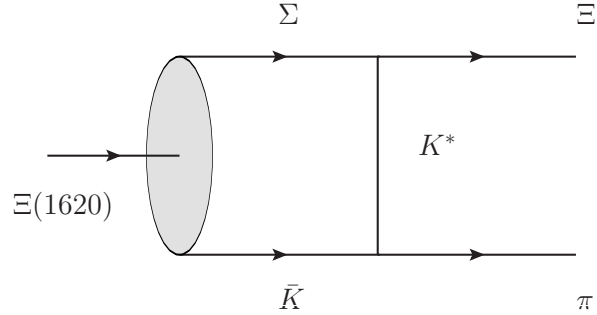}}
    \caption{Diagram contributing to the $\Xi(1620)\rightarrow \Xi^-\pi^+$ decay.}
  \label{decay}
\end{figure}

After obtaining the Bethe-Salpeter wave functions, we can calculate some physical properties of the molecular bound state which can be measured in experiments. One of the most important properties is the decay width. The bound state $\Xi(1620)$ can decay to $\Xi\pi$ via exchanging the $K^*$ meson as shown in Fig. \ref{decay}. There is no $K$ meson exchange contribution, as the spin-parity conservation forbids the vertex $KK\pi$. In the following we will write down the decay amplitude and calculate the decay width using the solution of the one-dimensional Bethe-Salpeter equation obtained in the previous section. The effective Lagrangian for the $BB K^*$ vertex is \cite{Nakayama:2006ty}
\begin{equation}
\mathcal{L}_{BB K^*}=-g_{BB K^*}\bar{B}\left(\gamma^\mu B K^*_\mu-\frac{\kappa_{BB K^*}}{m_N}\sigma^{\mu\nu}B\partial_\nu K^*_\mu\right).
\end{equation}
The Lagrangian for the vertex $K^*K\pi$ reads
\begin{equation}
\mathcal{L}_{K^*K\pi}=-ig_{K^*K\pi}K^{*\mu}(\boldsymbol{\pi}\partial_\mu-\partial\boldsymbol{\pi})\cdot\boldsymbol{\tau} K.
\end{equation}
where the coupling constants $g_{\Sigma\Xi K^*} = -3.52$, $\kappa_{\Sigma\Xi K^*} = 4.22$, and $g_{K^*K\pi} = -g_{\rho\pi\pi}/2$ with $g_{\rho\pi\pi} =6.1$ \cite{Nakayama:2006ty,He:2017aps}.

In the rest frame, we define $p'_1=(E'_1, p')$ and $p'_2=(E'_2,-p')$ ($p'$ is three-momentum) to be the momenta of $\Xi$ and $\pi$, respectively. The masses of $\Xi$ and $\pi$ are $m'_1$ and $m'_2$, respectively.  According to the kinematics in the rest frame of the two-body decay, one has
\begin{equation}
E'_1=\frac{M^2-m^{'2}_2+m^{'2}_1}{2M},\quad E'_2=\frac{M^2-m^{'2}_1+m^{'2}_2}{2M},
\end{equation}
\begin{equation}
|p'|=\frac{\sqrt{[M^2-(m^{'}_2+m^{'}_1)^2][M^2-(m^{'}_2-m^{'}_1)^2]}}{2M},
\end{equation}
and
\begin{equation}
d\Gamma=\frac{1}{32\pi^2}|\mathcal{M}|^2\frac{|p'|}{E^2}d\Omega,
\end{equation}
where $|p'|$ is the norm of the 3-momentum of either particle in the final state in the rest frame of the initial bound state and $\mathcal{M}$ is the Lorentz-invariant decay amplitude of the process.

According to the above interactions, the decay $\Xi(1620)\rightarrow\Xi^-\pi^+$ is shown in Fig. \ref{decay}. We can write down the amplitude as
\begin{equation}
\mathcal{M}=-\frac{g_{\Sigma\Xi K^*}g_{K^*K\pi}}{2}u_{\Xi}\gamma_\mu(p_2+p'_2)_\mu\Delta_{\mu\nu}(k,m_{K^*})u_{\Xi(1620)}\mathcal{F}^2(k)|_{k=q'-p}\chi_P(p),
\end{equation}
where $q'=\lambda_2q'_1-\lambda_1q'_2$ is not the relative momentum of particles in the final state, $\lambda_1$ and $\lambda_2$ are defined as $\lambda_i= m_i/(m_1+m_2)$, and $m_1$ and $m_2$ are the masses of the component particles of the bound states but not of the final state.

Then, we apply the numerical solution of the Bethe-Salpeter amplitude to calculate the decay width of $\Xi(1620)\rightarrow\Xi\pi$. The decay widths are 36.94 MeV and 9.35 MeV for the $\Lambda\bar{K}$ and $\Sigma\bar{K}$ bound stats, respectively.

\section{Summary}
\label{sec-summary}
In this paper, we applied the Bethe-Salpeter equations to study the possibilities that the is $\Xi(1620)$ is $s$-wave $\Lambda\bar{K}$ or $\Sigma\bar{K}$ bound states with the quantum numbers $J^P= 1/2^-$. Considering the interaction kernels based on $\omega$ and $\phi$ mesons exchange diagrams for the $\Lambda\bar{K}$ system and $\rho$, $\omega$, and $\phi$ mesons exchange diagrams for the $\Sigma\bar{K}$ system, we established the Bethe-Salpeter equations in the ladder and instantaneous approximations. Because the constituent particles and the exchanged particles in the $\Lambda\bar{K}$ and $\Sigma\bar{K}$ systems are not pointlike, we introduced a form factor including a cutoff $\Lambda$ which reflects the effects of the structure of these particles. Since $\Lambda$ is controlled by nonperturbative QCD and cannot be determined at present, we let it vary in a reasonable range within which we examined whether $\Lambda\bar{K}$ and $\Sigma\bar{K}$ bound states could be the $\Xi(1620)$ state by solving the Bethe-Salpeter equations. From our calculations, we found that the $\Xi(1620)$ state can be treated as the $\Lambda\bar{K}$ and $\Sigma\bar{K}$ bound states when $\Lambda$ = 1632 MeV and 1356 MeV, respectively.

Then, we applied the numerical solutions of the Bethe-Salpeter wave functions to calculate the decay widths of $\Xi(1620)\rightarrow\Xi\pi$ which are induced by $K^*$ exchange meson. We obtained that the decay widths are 36.94 MeV and 9.35 MeV for the $\Lambda\bar{K}$ and $\Sigma\bar{K}$ bound states, respectively. Comparing the magnitides of these two decay widths, it is obvious that the $\Xi(1620)$ has a larger contribution from the $\Lambda\bar{K}$ system than the $\Sigma\bar{K}$ system. The same conclusion was also found in Ref. \cite{Ramos:2002xh} from the chiral perturbation theory. Clearly, more theoretial and experimental efforts will be needed to fully understand the nature of the one-star $\Xi(1620)$.

\acknowledgments
This work was supported by National Natural Science Foundation of China (Projects No. 11575023, No. 11775024, No. 11605150), Natural Science Foundation of Shandong Province of China No. ZR2016AQ01 and K.C.Wong Magna Fund in Ningbo University.

\end{document}